\documentstyle[aps]{revtex}

\newcommand{\be}{\begin{equation}}
\newcommand{\ee}{\end{equation}}
\newcommand{\vv}[1]{\mbox{\boldmath $#1$}}
\newcommand{\vtimes}{\vv{\times}}

\newcommand{\Hion}{H_{\rm ion}}                % ion Hamiltonian
\newcommand{\Hat}{H_{\rm at}}                  % atom Hamiltonian
\newcommand{\He}{H_{\rm e}}                    % electron Hamiltonian

\newcommand{\Heff}{H_{\rm eff}}                % effective one-particle Hamiltonian
\newcommand{\me}{m_{\rm e}}                    % electron mass
\newcommand{\aB}{a_0}                          % Bohr radius
\newcommand{\am}{a_{\rm m}}                    % magnetic length
\newcommand{\muB}{\mu_{\rm B}}                 % Bohr magneton
\newcommand{\E}{\varepsilon}                   % shifted energy
\newcommand{\Ry}{\rm Ry}
\begin{document}
\title{Bound states of negatively charged ions induced by a magnetic field}
\author{Victor G. Bezchastnov, Peter Schmelcher, and Lorenz S. Cederbaum}
\address{Theoretische Chemie, Physikalisch-Chemisches Institut,
Universit\"at Heidelberg, INF 229, D-69120 Heidelberg, Federal Republic of Germany}
\date{\today}
\maketitle
\begin{abstract}
We analyse the bound states of negatively charged ions which were predicted to 
exist because of the presence of a magnetic field by Avron {\it et al} \cite{Avron}. 
We confirm that the 
number of such states is infinite in the approximation of an infinitely heavy nucleus 
and provide insight into the underlying physical picture by means of a combined 
adiabatic and perturbation theoretical approach. 
We also calculate the corresponding binding energies which are 
qualitatively different for the 
states with vanishing and non-vanishing angular momentum. 
An outlook on the case of including center of mass effects is presented. 
\end{abstract}
\section{Introduction}
The behaviour and properties of negative ions 
became during the past years a branch of intense research.
There exists an enormous interest in the electronic structure and
dynamics of negative ions, both 
from the theoretical as well as experimental point of view 
(see, e.g., the review \cite{Scheller} and the references therein). 
According to our present knowledge about atomic ions it is most likely 
that singly charged negative ions possess in the absence of a magnetic field only
one stable ground-state configuration. For example for the 
H$^-$ ion, this state is the $1\,^{1}S$ electronic state, 
and a rigorous proof that this the only possible bound state was given in \cite{Hill}. 
Also, some atoms, like for example 
%Be, N, Ne, Mg, Ar, Ca, Sc, Mn, Zn, Kr, Sr, Cd, Xe, Ba, Hf, Hg, Rn 
Be, N, Ne, Mg, Ar do not possess any stable 
negative ion state (see, e.g., \cite{Smirnov,Andersen}). 
On the other hand, one can expect that in the presence of a magnetic field a
lot of new discrete energy states of negative ions can appear. 
This expectation is based on the 
statement that for any negatively 
charged ion the number of discrete energy states is infinite in the presence 
of a magnetic field \cite{Avron}. 
However, this statement was formulated as the conclusion 
of a formal mathematical treatment which does not provide a 
transparent physical picture of the appearance of the infinite sequence of bound 
states nor does it estimate the corresponding energies. Although a wide set 
of works was focused on the influence of the magnetic field on low-lying ion 
states (e.g., \cite{Henry,Surmelian,Larsen,Park,Vincke,Larsen1,Bylicki}), 
there is no systematic treatment of 
the highly excited anions predicted in \cite{Avron}. 
In the present paper we develop a physical approach showing 
transparently from which quantum mechanical grounds the infinite 
sequence of bound anion states appears in the presence of a magnetic field.
In Section~2, we first reduce the problem to a 
one-particle, and then to a one-dimensional one, where to analyse the bonding one 
has to consider the motion of the external electron along the magnetic field in an 
effective potential depending on the quantum state. We confirm that, in the 
approximation of the static (infinitely heavy) nucleus, the number of the 
bound states is infinite, and derive estimates 
for the corresponding binding energies valid for arbitrary ions. 
In Section~3, we apply these estimates to the ion H$^-$ and verify them treating the 
one-dimensional motion of the external electron numerically. We also discuss the bound 
states of an exotic ion formed by attaching the muon to the hydrogen atom. Besides, 
we analyse the importance of the non-adiabatic coupling between the states of the external 
and atomic (core) electrons. A general discussion and concluding remarks are given in 
Section~4.
\section{General}
It is well-known (see, e.g., \cite{Smirnov,Massey}) that the electron affinities for the 
negative ions are usually much smaller 
than the binding energies of the corresponding neutrals. This fact implies that the extra 
electron is weakly bound to the atom and its behaviour and properties strongly differ 
from that of internal 
(core) electrons. For the H$^--$ion, the binding of the extra electron is 
exclusively due to correlation between the two electrons. 
In the following we 
will study the highly excited states of the negative ions which appear in the presence of 
a magnetic field and which, as we will estimate, correspond to very small electron affinities. 
We therefore use the approximation of a weakly bound external electron neglecting 
its exchange interaction with the core electrons. To exploit such a model, 
it is convenient to split 
the total Hamiltonian for a singly negatively charged ion into three terms,
\be
\Hion = \Hat + \He + W.
\label{Hion}
\ee
Here the first term,
\be
\Hat = \frac{1}{2\me} \sum_{a=1}^Z\left[ \vv{p}_a + \frac{e}{c}\,\vv{A}_a \right]^2 + U 
     + 2 \mu_B \vv{B} \vv{\Sigma}
\label{Hat}
\ee
describes a neutral atom with infinitely heavy nucleus in a magnetic field $\vv{B}$ 
with the vector potential $\vv{A}_a = \vv{A}(\vv{r}_a)$.
Summation is carried out over all the atomic (core) electrons labelled with 
the subscript ``$a$'',
$Z$ is the nucleus charge number, 
$\me$ and $-e=-|e|<0$ are the electron mass and charge.  
The potential energy $U$ 
includes the Coulomb energies of the interaction of the core electrons with each other 
and with the nucleus. 
$\vv{\Sigma}$
%=\sum_{a=1}^Z\vv{s}_a$ 
is the total spin of the atom, and 
$\muB=e\hbar/(2\me c)$ is the Bohr magneton. 
The second term in (\ref{Hion}),
\be
\He = \frac{1}{2\me} \left[ \vv{p} + \frac{e}{c}\,\vv{A} \right]^2 
    + 2 \mu_B \vv{B} \vv{\sigma},
\ee
corresponds to an extra (with respect to the atom) electron, 
$\vv{\sigma}$ is the electron spin.
%moving in a magnetic field. 
The last term,
\be
W = \sum_{a=1}^Z\frac{e^2}{|\vv{r}-\vv{r}_a|} - \frac{Ze^2}{r},
\label{W}
\ee
describes the Coulomb coupling of the extra electron to the atom. 

When the extra electron is weakly bound to the atom, the 
character of its motion differs strongly 
from that of the core electrons. In particular, the external electron can be assumed to move much 
slower than the core ones. 
In this case, the ion states can be successfully described in terms of 
a quasimolecular approach, and the problem of binding can even be reduced to a one-particle one. 
Below we briefly describe this approach.
\subsection{Quasimolecular approach to the problem of binding}
Let us consider the Hamiltonian
\be
\Hat' = \Hat + W.
\label{H'}
\ee 
It does not include the kinetic energy of the external electron, 
therefore the latter can be considered as static in space, so 
$\Hat'$ describes the motion of the core electrons only. 
Since the potential energy term of $\Hat'$ depends parametrically on the position $\vv{r}$ 
of the external electron, the energy spectrum of this Hamiltonian as well as the corresponding 
eigenfunctions also depend on $\vv{r}$. Let us denote them by 
\be
E'_i(\vv{r}) = \langle i | \Hat' | i \rangle
\ee
and    
$\phi_i(\vv{r}_1,\vv{r}_2,\ldots,\vv{r}_Z;\vv{r})$, 
respectively, where the index $i$ labels the eigenstates of 
the Hamiltonian $\Hat'$ with some choice of $\vv{r}$. Since at any $\vv{r}$ the states
$|i\rangle$ compose a complete basis set, an eigenfunction of the total Hamiltonian (\ref{Hion}) 
can be presented as
\be
\Psi(\vv{r}_1,\vv{r}_2,\ldots,\vv{r}_Z;\vv{r}) = 
\sum_{i'} \psi_{i'}(\vv{r}) \phi_{i'}(\vv{r}_1,\vv{r}_2,\ldots,\vv{r}_Z;\vv{r}),
\label{exp}
\ee
with the expansion coefficients $\psi_{k'}(\vv{r})$ depending on the position of the external 
electron. 
It should be noted that this wave function is not completely antisymmetric with 
respect to all the ion electrons, only its ``atomic part'', $\phi_{i'}$ can be assumed to be 
properly antisymmetric with respect to the core electrons. Complete antisymmetrisation would 
significantly complicate the further consideration. On the other hand, the main yield of such a 
complication would be taking into account exchange interaction between the external and core 
electrons which is definitely negligible for states for which the external electron is 
localized far from the atom. 
Notice also that the form (\ref{exp}) of the ion wave function is exact if another 
charged particle different from the electron (like for example a muon) is attached to the atom.  

Substituting the wave function (\ref{exp}) into the Schr\"odinger equation with the Hamiltonian 
(\ref{Hion}), subsequently multiplying from the left-hand side by 
$\phi^\ast_i$ and integrating over the positions of the core electrons 
we will arrive at 
the set of coupled equations
\be
\left[ \He + E'_i(\vv{r}) + h_{ii}(\vv{r}) - E_{\rm tot} \right] \psi_i(\vv{r}) = 
- \sum_{i' \neq i} h_{i\,i'} \, \psi_{i'}(\vv{r}).
\label{ceqs}
\ee
Here $E_{\rm tot}$ is the total eigenenergy of the system, and the matrix elements 
$h_{i\,i'}$ are given by 
the following equation
\be
h_{i\,i'} = \frac{1}{2\me} \langle \phi^\ast_i | p^2 \phi_{i'} \rangle
          + \frac{e \vv{B}}{2 \me c} \langle \phi^\ast_i | \vv{l} \, \phi_{i'} \rangle
          + \frac{1}{\me} \langle \phi^\ast_i | \vv{p} \, \phi_{i'} \rangle \vv{p}~,
\label{h-def}
\ee
where $\vv{l} = \vv{r} \times \vv{p}$ is the angular momentum of the external electron. Notice that 
if both $i$ and $i'$ relate to the ground state of the atom, $|0\rangle$,
then the last two terms in Eq.~(\ref{h-def}) equal 
zero while the first term can be transformed as follows
\be
h_{0\,0} = \frac{\hbar^2}{2\me} 
           \int {\rm d} \vv{r}_1 \int {\rm d} \vv{r}_2 \ldots \int {\rm d} \vv{r}_Z 
           \left| \frac{\partial}{\partial \vv{r}} 
                  \phi(\vv{r}_1,\vv{r}_2,\ldots,\vv{r}_Z;\vv{r}) \right|^2~. 
\label{h_00}
\ee

In the following we restrict ourself by considering the attachment of the extra electron 
to the neutral atom in the ground state. In this case, 
the right-hand side of Eq.~(\ref{ceqs}) is associated with the coupling between the ground and 
excited states of the core electrons due to the motion of the external electron. 
Assuming the ground state of the neutral atom to be energetically well-separated 
from the excited states this coupling can be neglected. 
Then Eq.~(\ref{ceqs}) becomes essentially a one-particle Schr\"odinger equation,
\be
\left[ \He + V_0(\vv{r}) + h_{00}(\vv{r}) - E_0 \right] \psi_0(\vv{r}) = 0, 
\label{one-particle_corr}
\ee
where, for the sake of convenience, we have shifted the energies by the ground state energy of 
the isolated atom, $E^{(0)}_0 = E'_0(r\to\infty)$, and introduced 
\be
E_0 = E_{\rm tot} - E^{(0)}_0
\label{E_i}
\ee
and 
\be
V_0(\vv{r}) = E'_0(\vv{r}) - E_0^{(0)}.
\label{V_i}
\ee 
Equation (\ref{one-particle_corr}) describes the 
motion of the external electron in the magnetic field and the potential which 
consists of two parts: the static term, $V_0(\vv{r})$, and the non-adiabatic (dynamic) 
correction, $h_{00}(\vv{r})$. The latter, as we shall directly estimate for the H$^--$ion, 
decreases with increasing $r$ faster than $V(\vv{r})$ and can be neglected at distances 
exceeding the atomic size. Therefore, to analyse the binding mechanism for the excited 
ion states one has to solve the Schr\"odinger equation
\be
\left[ \He + V(\vv{r}) - E \right] \psi(\vv{r}) = 0, 
\label{one-particle}
\ee
where for brevity we have omitted the index ``0'' related to the ground state 
of the core electrons. 
 
At distances of the extra 
electron from the atom strongly exceeding the atom size, one can 
consider the operator (\ref{W}) as a 
perturbation of the atom and evaluate it by the 
multipole expansion,
\be
W = -e \left( \frac{\vv{D}\vv{n}}{r^2} 
            - \frac{Q_{\alpha \beta} n_\alpha n_\beta}{2r^3} + \ldots \right)~,
\label{mult}
\ee
where $\vv{n}=\vv{r}/r$ is the unit vector in the direction of $\vv{r}$ and the  
indices $\alpha$ and 
$\beta$ run over the Cartesian coordinates.
\be
\vv{D} = -e \sum_{a=1}^Z \vv{r}_a
\ee 
and 
\be
Q_{\alpha \beta} = -e \sum_{a=1}^Z 
                   \left( r_a^2 \delta_{\alpha \beta} - 3 x_{a\alpha} x_{a\beta} \right)
\ee
are the operators of the dipole and quadrupole momenta of the atom, respectively, 
$x_{a\alpha}$ and $x_{a\beta}$ denote the components of $\vv{r}_a$, 
$\vv{r}_a=(x_{a1},x_{a2},x_{a3})$. 
Then the potential in Eq.~(\ref{one-particle}) can be approximated as
\be
V(\vv{r}) = -\frac{e \langle \vv{D} \rangle \vv{n}}{r^2} 
              +\frac{e \langle Q_{\alpha \beta} \rangle n_\alpha n_\beta}{2r^3} 
              -\frac{e^2 \kappa_{\alpha \beta} n_\alpha n_\beta}{2r^4}~,
\label{V_large_r}
\ee
where $\langle \vv{D} \rangle$ and $\langle Q_{\alpha \beta} \rangle$ are the mean values of 
the dipole and quadrupole momenta, respectively, of the unperturbed atom 
described by the Hamiltonian (\ref{Hat}), 
and $\kappa_{\alpha \beta}$ is the polarizability 
of the atom in an electric field. 
The first two terms in (\ref{V_large_r}) represent 
the first-order perturbation corrections. The last term in (\ref{V_large_r}) 
is the second-order correction, it corresponds to the dipole term in (\ref{mult}) 
treated as the perturbation of the atom by the electric field 
$\vv{\cal E} = -e \vv{n} / r^2$.

Expression (\ref{V_large_r}) can be further simplified if we take into account that the 
potential (\ref{V_i}) has its symmetry axes directed along the magnetic field. 
A simple argumentation of this statement is the following. 
In the presence of the magnetic field it is natural to assume that the 
axis of quantization for the atom is directed along $\vv{B}$. Then the 
potential (\ref{V_i}) depends parametrically on the two vectors, $\vv{B}$ and $\vv{r}$. 
Since the potential is scalar it must depend only on scalars which can be constructed from 
these vectors. These scalars are $B^2$, $r^2$ and $\vv{B}\vv{r}$. This means that  
if $\vv{B}$ is directed along the $z-$axes, then $V(\vv{r}) = V(r_\perp,z)$. 
As a result the non-zero components of the mean dipole and quadrupole momenta 
of the neutral atom can only be $\langle D_z \rangle$,  
$\langle Q_{xx} \rangle = \langle Q_{yy} \rangle$ and $\langle Q_{zz} \rangle$, 
respectively. Also, the polarizability takes the diagonal form with the components
$\kappa_{xx} = \kappa_{yy} = \kappa_\perp$ and 
$\kappa_{zz} = \kappa_\parallel$. 
Furthermore, since the $z-$parity of the 
core electrons is the integral of motion when they are not perturbed by the interaction with 
the external electron, we have $\langle D_z \rangle = 0$. 
Then Eq.~(\ref{V_large_r}) can be transformed as follows:
\be
V(r_\perp,z) = \frac{e^2 \lambda (2z^2 -r_\perp^2)}{2r^5} 
             - \left( \kappa_\perp \sin^2\vartheta 
                    + \kappa_\parallel \cos^2\vartheta \right) \frac{e^2}{2r^4}~, 
\label{V_large_r1}
\ee
where $\vartheta$ is the angle between $\vv{n}$ and $\vv{B}$ and the coefficient
$\lambda$ can be expressed
in terms of the mean values of the squares of the longitudinal and transverse 
coordinates of the core electrons of the unperturbed atom, 
\be
\lambda = \sum_{a=1}^Z \left[ 2 \langle z_a^2 \rangle 
                           - \langle r_{a\perp}^2 \rangle \right]~.
\ee
\subsection{Binding in a magnetic field as 1D problem}
The Hamiltonian that determines the one-particle Schr\"odinger equation (\ref{one-particle})
explicitly reads
\be
\Heff = \frac{p_z^2}{2\me} 
      + \frac{\pi_\perp^2}{2\me}  
      + V(r_\perp,z),
\label{Heff}
\ee
where
\be
\vv{\pi_\perp} = \vv{p}_\perp + \frac{e}{2c}\,\vv{B}\vtimes\vv{r}_\perp
\ee
is the transverse kinetic momentum of the electron.  
We have introduced the symmetric gauge of the vector potential, 
$\vv{A} = (1/2) \vv{B} \vtimes \vv{r}$, being the most appropriate one because of the axial 
symmetry of the potential $V$. We also have omitted the spin part of the Hamiltonian 
which determines the trivial shift of the ion energy spectrum and does not affect the 
binding energies.

Since the potential in the Hamiltonian (\ref{Heff}) 
decreases rapidly with increasing distance of the extra electron from 
the atom, even quite a weak magnetic field can influence the transverse motion of 
the extra electron to a much larger extent than the atomic potential. In this case the 
eigenfunction of the Hamiltonian (\ref{Heff}) can be efficiently expanded in terms of the 
the Landau states, $\langle \vv{r}_\perp | n, s \rangle$, of the motion of the electron 
across the magnetic field. These states are the common eigenstates of the operators 
$\pi_\perp^2$ and $l_z$, where $\vv{l}=\vv{r}\vtimes\vv{p}$ is the electron angular 
momentum, and labeled with the two quantum numbers, $n$ and $s$, that determine the 
corresponding eigenvalues,
\begin{eqnarray}
\pi_\perp^2 = (\hbar/\am)^2 (2n+1)~, &\;\;& n=0,1,2,\ldots,
\nonumber \\
l_z = -\hbar s~, &\;\;& s=-n,-n+1,-n+2,\ldots,
\label{eigenvalues}
\end{eqnarray}
$\am = \sqrt{c\hbar/(eB)}$ is the magnetic length. The first eigenvalue determines 
the Landau energy spectrum
\be
E^{\rm Lan}_n = \frac{\pi_\perp^2}{2\me} 
              = \frac{\hbar e B}{\me c} \left( n + \frac{1}{2} \right)~.
\label{Landau_energies}
\ee
Hence  $n$ is called the electron Landau level number.

Because of the axial symmetry of the atomic potential the longitudinal component of the 
electron angular momentum is an integral of motion for the Hamiltonian (\ref{Heff}), 
$[l_z,\Heff]=0$, and the quantum number $s$ can be used to label the eigenstates of $\Heff$, 
i.e. $E=E_s$. The expansion of the corresponding eigenfunctions over the Landau states 
thus involves only different Landau level numbers and reads 
\be
\psi_s(\vv{r}) = \sum_n g_{ns}(z) \langle \vv{r}_\perp | n, s \rangle. 
\label{Landau_expansion}
\ee
The expansion coefficients $g_{ns}(z)$ and the energy $E$ can be found by solving the 
system of coupled equations,
\be
\left[ \frac{p_z^2}{2\me} + V^{(s)}_{nn}(z) + E^{\rm Lan}_n - E_s \right] g_{ns}(z) = 
- \sum_{n' \neq n} V^{(s)}_{nn'}(z) g_{n's}(z)~, \;\;\; n=0,1,2,\ldots,
\label{ceqs1}
\ee
where 
\be
V^{(s)}_{nn'}(z) = \langle n,s | V(r_\perp,z) | n',s \rangle
\ee
are the effective longitudinal potentials obtained as the matrix elements of the 
atomic potential $V(r_\perp,z)$ within the space of the Landau states of the 
external electron. 

The quantity $E^{\rm Lan}_0 = \hbar \omega_B /2$, where $\omega_B = eB/(\me c)$ is the 
electron cyclotron frequency, determines the continuum threshold for the states of the 
external electron. The states of this electron belong to different 
manifolds associated with the different Landau energies 
$E^{\rm Lan}_n$, $n=0,1,2,\ldots$. 
The bound states can be classified by the 
numbers $(s,n,\nu)$ where $\nu$ equals the number of nodes of the function $g_{ns}(z)$ for 
the leading term in the expansion (\ref{Landau_expansion}) - such a classification 
is similar to that used in treating hydrogen-like atoms in a strong magnetic field, 
see, e.g., \cite{Ruder}. 

We can expect (and our results for the ion H$^-$ 
confirm this), that the binding energies for the ionic states induced by the presence of the 
magnetic field are small compared to the electron cyclotron energy, $\hbar \omega_B$. 
Therefore, all such states associated with the manifolds $n=1,2,\ldots$ lie in the continuum, 
i.e. are not bound. We thus will focus on the states related to the ground 
Landau manifold, $n=0$. While treating them we can further neglect the coupling of 
the ground Landau manifold to the higher ones, e.g., omit the sum on the right-hand side 
of Eq.~(\ref{ceqs1}). This approach yields the 1D Schr\"odinger equation,
\be
\left[ \frac{p_z^2}{2\me} + V_s(z) - \E_s \right] g_s(z) = 0, 
\label{1D}
\ee
which describes the motion of the external electron along the magnetic field. In this equation,
we have shifted the energy by the zero-point Landau energy, 
\be
\E_s = E_s - E^{\rm Lan}_0~.
\label{shifted_energy}
\ee
Then, 
when $\E_s$ is negative, the value of $-\E_s$ is the binding energy of the external electron. 
Also, for brevity we have omitted the index $n=0$ and define the effective longitudinal 
potential by $V_s(z) = V^{(s)}_{00}(z)$. 
In order to get an idea of the binding properties of this potential one can use  
the weak coupling one-dimensional theory (see, e.g., \cite{Simon}). This theory says that 
if the potential 
%is such that $V_s(z) \to 0$ for 
vanishes at $|z| \to \infty$ and the integral  
\be
I_s = \int_{-\infty}^\infty V_s(z) {\rm d}z
\label{integral}
\ee
is negative than 
there is at least one bound state of finite longitudinal motion, 
and the estimate of the corresponding binding energy for that state is
%\be
%-\E_s = \frac{\me}{2\hbar^2} 
%        \left( \int_{-\infty}^{+\infty} V_s(z) {\rm d}z \right)^2~.
%\label{bind}
%\ee
\be
-\E_s = \me I_s^2 /(2\hbar^2)~.
\label{bind}
\ee
As we shall see, the condition $I_s < 0$ holds 
for infinitely many possible quantum numbers $s$ and thus the number of bound 
states associated with different values of $s$ is infinite, 
in accordance with the conclusion of \cite{Avron}. For each $s$, the estimate (\ref{bind}) 
gives the binding energy for the state for which the longitudinal part of the wave function 
of the external electron has no nodes, e.g. $\nu=0$.

To exploit the 
%last two equations 
latter analytical approach 
let us introduce the transverse probability density for the states 
related to the ground Landau manifold
\be
\rho_s(r_\perp) = \int_0^{2\pi}{\rm d}\varphi \, | \langle \vv{r}_\perp |n=0,s \rangle |^2   
                = \frac{1}{s!\,\am^2} 
                  \left( \frac{r_\perp^2}{2\am^2} \right)^s 
                  \exp \left( - \frac{r_\perp^2}{2\am^2} \right)~,
\label{rho_s}
\ee
where the integration is done over the electron azimuthal angle. 
%Now we can rewrite the condition (\ref{cond}) 
%and the estimate of the corresponding binding energy 
In terms of the density (\ref{rho_s}) 
and the atomic potential $V(r_\perp,z)$ 
%as
we have 
\be
I_s = \int_0^\infty r_\perp {\rm d}r_\perp \, \rho_s(r_\perp) 
      \int_{-\infty}^\infty {\rm d}z \, V(r_\perp,z)~.
\label{integral1}
\ee
%-\E_s = \frac{\me}{2\hbar^2} \left[
%        \int_0^\infty r_\perp {\rm d}r_\perp \, \rho_s(r_\perp)        
%        \int_{-\infty}^{+\infty} {\rm d}z \, V(r_\perp,z) \right]^2~,
%\label{bind1}
%\ee
%respectively. 
Let us now adopt an approximation of the atomic potential 
given by Eq.~(\ref {V_large_r1}) for large distances. It is easy to check 
%(e.g., by the integration by parts) 
that at any $r_\perp \neq 0$
\be
  \int_0^\infty \frac{\left(2z^2-r_\perp^2\right){\rm d}z}
                     {\left( r_\perp^2 + z^2 \right)^{5/2}} = 0
%= \int_0^\infty \frac{3z^2{\rm d}z}
%                       {\left( r_\perp^2 + z^2 \right)^{5/2}}  
%- \int_0^\infty \frac{{\rm d}z}
%                     {\left( r_\perp^2 + z^2 \right)^{3/2}} = 0
\ee
and thus the first, quadrupole, term in (\ref {V_large_r1}) 
%counts neither in Eq.~(\ref{cond1}) nor in Eq.~(\ref{bind1}). 
does not contribute into $I_s$. 
This implies that the polarization part of the atomic potential 
plays a key role in binding the external electron. 
Let us further assume that the influence of the magnetic 
field on the atomic polarizability is negligible. As it is known \cite{Bonin}, 
in the absence of the magnetic field 
an atom in the ground state of the zero total angular momentum has an isotropic 
polarizability, $\kappa$, i.e. we assume 
$\kappa_\perp \approx \kappa_\parallel \approx \kappa$. 
Then the polarization term 
%of the atomic potential 
in 
(\ref {V_large_r1}) is net attractive 
and the condition $I_s<0$ is fulfilled. 
Straightforward integration in Eq.~(\ref{integral1}) 
and use of the relation (\ref{bind}) 
gives
\be
-\E_s = \frac{\pi^2 \kappa^2}{2^7 (s!)^2} \Gamma^2 \left( s - \frac{1}{2} \right) 
        {\rm Ry} \, \gamma^3~,
\label{s>0}
\ee 
where $\Gamma$ is the complete gamma-function, ${\rm Ry} = \me e^4 / (2 \hbar^2)$, 
$\gamma = \am^2/\aB^2 = B/B_0$, and $B_0 = 2.35 \times 10^5$~T. 
This estimate is not appropriate for $s=0$ because of the divergence 
of the integrations at the origin $r_\perp = z = 0$ for the potential 
(\ref {V_large_r1}). However, for $s>0$ above value of the binding energy is a good 
approximation for $\gamma \ll 1$ (laboratory magnetic fields) 
since the main contribution to the integral comes from distances sufficiently 
larger than $\aB$, the natural measure of the size of the neutral atom. 
Using atomic units 
($2\Ry$ for the energy and $B_0$ for the magnetic field strength) 
we rewrite Eq.~(\ref{s>0}) in a recurrent way,
\be
-\E_s = 0.1211 \kappa^2 B^3 \delta_s^2~,\;\;\;s=1,2,\ldots,
\label{Es}
\ee
where 
\begin{eqnarray}
\delta_1 &=& 1~,
\nonumber \\
\delta_s &=& [1-(1.5/s)] \delta_{s-1}~, \;\;\; s = 2,3,\ldots~. 
\end{eqnarray}
One should notice that the scaling behaviour of the binding energies with 
the magnetic field strength, $-\E_s \propto B^3$, coincides with that predicted in 
\cite{Avron}. Also, at large $s$ we have $\delta_s \approx s^{-3/2}$ and thus 
obtain the behaviour of the binding 
energies for large $s$, $-\E_s \propto s^{-3}$. Such a sharp decrease of the binding 
energy with the quantum number reflects the fact that in the plane perpendicular to the 
magnetic field the external electron follows the Landau orbit with the expectation 
value for the radius squared $\langle r_\perp^2 \rangle = 2(s+1)\am^2$, 
and at large $s$ the transverse probability density (\ref{rho_s}) becomes a peaked 
function of $r_\perp^2$ centered around 
$r_\perp^2 = \langle r_\perp^2 \rangle$. Therefore, as $s$ increases 
the external electron is bound to the atom at progressively 
larger distances from it.

We can also establish the behaviour of the binding energy with the magnetic 
field strength for the state $s=0$. In this case, contrary to the limit of large $s$, 
the transverse probability density varies very slowly with $r_\perp$ and the scale on which it 
reduces significantly from its maximal value at $r_\perp=0$ is the magnetic length, $\am$, 
which strongly exceeds the atomic potential scale, $\aB$. Therefore, when 
%estimating the binding energy from Eq.~(\ref{bind1}) 
calculating the integral (\ref{integral1}) 
we can replace the transverse probability density by its 
maximal value, $\rho_s(0) = 1/\am^2$. 
In this way we obtain
\be
-\E_0 = \frac{1}{2} 
        \left[\int_0^\infty r_\perp {\rm d}r_\perp         
              \int_{-\infty}^{+\infty} {\rm d}z \, V(r_\perp,z) \right]^2
        B^2~,
\label{s=0}
\ee
where we have used atomic units again as in the remaining part of our work. 
%the units $2\Ry$, $\aB$ and $B_0$ to measure the energy, atomic potential, 
%transverse and longitudinal distances and the magnetic field strength, respectively. 
We remark that the domain of 
the neutral atomic core contributes quite significantly 
to the integrations in (\ref{s=0}) and thus the use of the 
quasimolecular (adiabatic) picture to describe the motion of the extra electron becomes a 
crude approximation. 
However, a more accurate treatment is expected to change only 
the numerical coefficient in Eq.~(\ref{s=0}). 
The scaling behaviour of the binding energy with respect to the field strength is 
expected to be the same, i.e. $-\E_0 \propto B^2$. 
\section{Binding energies of the H$^-$ and H$\mu^-$ ions}
The approach developed in the previous Section can be directly applied to the 
H$^--$ion. With a good accuracy we can neglect the influence of the magnetic field 
on the ground state of the core electron, and use the well-known value 
(see, e.g., \cite{Miller}), 
$\kappa=9/2$, for the polarizability of such a configuration. Then the estimate 
(\ref{Es}) gives
\be 
-\E_s = 2.453 B^3 \delta_s^2~.
\label{E_s}
\ee

More reliable results for the binding energies for the states of the electron attached to the 
hydrogen atom which are induced by the presence of the magnetic field, including the state 
$s=0$, can be obtained if one knows the atomic potential $V$. When the influence of 
the magnetic field on the hydrogen atom is neglected this potential is spherically symmetric, 
$V=V(r)$, and can be found in the literature. We use the results of Walles, Herman and Milnes 
\cite{Walles} who rigorously calculated the energy levels of an electron moving in the field of two 
fixed, at distance $r$ from each other, 
charges $+e$ and $-e$ using the separability of the problem in confocal elliptic 
coordinates. Their electronic ground state energy, $E_{\rm WHM}$, 
is directly related to our potential $V(r)$, 
\be
V(r) = E_{\rm WHM}(r) - \frac{1}{r} + 0.5~,
\label{connection}
\ee
when adding the Coulomb interaction energy between the two static charges 
and the hydrogenic ground-state binding energy (cf. Eq.~(\ref{V_i})). 
The energy $E_{\rm WHM}$ was tabulated in the wide range of $r$, 
$0.840380 < r < 30.0$. In Figure~1, we show the corresponding values of $V$ by dots and use 
the spline interpolation to determine the value of the potential between them (solid line).
At larger $r$ the potential can be evaluated by its polarization tail,
\be
V(r) = -\frac{2.25}{r^4}~,
\label{large_r}
\ee
while at the lower $r$ one can put $E_{\rm WHM}=0$ in Eq.~(\ref{connection}) and get 
\be
V(r) = - \frac{1}{r} + 0.5~.
\label{small_r}
\ee
As Figure~1 demonstrates, these extrapolations match very good the data from \cite{Walles}.

With the well-defined potential $V(r)$ we can find the value of the coefficient in the estimate 
of the binding energy for the state $s=0$. Numerical integration of the potential in 
Eq.~(\ref{s=0}), with the natural replacement 
$2\pi r_\perp {\rm d} r_\perp {\rm d} z \to 4\pi r^2 {\rm d} r$, yields 
\be
-\E_0 = 6.31 B^2~.
\label{E_0}
\ee

To verify the estimates (\ref{E_s}) and (\ref{E_0}) we have also solved numerically 
the Schr\"odinger equation (\ref{1D}), calculating 
the effective potentials $V_s(z)$ numerically from the potential $V(r)$ 
presented in Figure~1. First of all, we were searching for the states with no 
nodes of the longitudinal wave function of the external electron ($\nu=0-$states). 
The results for the corresponding binding energies, for different numbers of $s$ 
and field strengths, are shown in Figure~2 by open circles. Solid lines represent the 
estimates (\ref{E_s}) and (\ref{E_0}), and we can conclude that they 
are in fairly good agreement with the numerical results. We have also performed the 
search for the bound states with higher longitudinal excitations, $\nu=1,2,\ldots$, but 
found no one, for the magnetic field strengths $10^{-6} < B < 10^{-4}$. This indicates 
that the effective potential $V_s(z)$ is so weak that for each $s$ it can bind 
the external electron only in the $\nu=0-$state.

Finally, to control the validity of the quasimolecular approach, we have calculated 
the quantity $h(r)=h_{00}(r)$, the non-adiabatic correction to the potential $V(r)$, 
determined by Eq.~(\ref{h_00}). For this purpose, the wave functions of the 
electron moving in the field of two fixed charges were computed 
via a numerical two-dimensional grid method which was invented by M.V.~Ivanov 
(for details of the method as well as the 
description of the corresponding code ATMOLMESH we refer the reader to \cite{Ivanov}). 
The corresponding results are shown in Figure~1 by open circles. 
To confirm that the numerical results 
reproduce correctly the scaling behaviour of $h$ with $r$, 
we also give the perturbation estimate, 
\be
h(r) = \frac{8}{r^6}~,
\label{h}
\ee
whose derivation is outlined in the Appendix. 
With increasing distances exceeding the size of the neutral hydrogen atom, the non-adiabatic 
correction becomes smaller and smaller compared to $|V(r)|$. 
One should notice 
that at small $r$ the non-adiabatic correction exceeds the value of $|V(r)|$. 
This implies that the non-adiabatic coupling effects are quite  
significant for the state $s=0$. However, since the corresponding binding energy 
obtained neglecting these effects is very small, we believe that a more accurate approach 
would only change the numerical coefficient in Eq.~(\ref{E_0}) but not the 
scaling low, $-\E_0 \propto B^2$.

The approach developed in this paper can also be directly applied to the analysis of 
highly excited exotic anions, for example to the ion formed by attaching the muon 
to the hydrogen atom. Due to the fact that the shell of the H$\mu^--$ion is 
formed by two different particles (the electron and muon) there is no need for 
antisimmetrization (exchange) and the ansatz (\ref{exp}) for the wave function of the ion 
becomes exact. Moreover, 
since the muon is much heavier than the electron (the muon-to-electron mass ratio is 
$m_\mu/\me \approx 207$), the binding energies for the muonic ion are larger. 
Replacement of the electron mass by the muon mass in Eq.~(\ref{bind}) leads to 
the following estimates
\begin{eqnarray}
-\E_0 &=& 1306 B^2~,\;\;\;s=0~,
\nonumber \\
-\E_s &=& 508 B^3 \delta_s^2~,\;\;\;s=1,2,\ldots~.
\end{eqnarray}
In Figure~2, the corresponding energies are plotted as dashed lines. Evidently, 
the non-adiabatic correction to the one-particle potential for the H$\mu^-$ ion 
is $m_\mu/\me$ times smaller than that for the ion H$^-$ (see Figure~1), 
which makes the estimate for $s=0$ more reliable in the case of the H$\mu^-$ ion.

\section{Summary and Outlook}
In this paper, we have studied the 
%highly excited 
states of the negative atomic ions which are 
induced by the presence of a magnetic field. 
Apart from the neglect of the exchange interaction between the external and 
atomic (core) electrons the equations we start with are exact. 
Performing an analysis of the weakly bound states of the external electron, 
it is possible to reduce the problem to an effective one-particle by neglecting 
the non-adiabatic coupling terms. In the presence of the magnetic field 
which determines the transverse motion of the external electron further 
simplifications are possible which reduce the question of the binding mechanism 
to a one-dimensional Schr\"odinger equation 
for the motion of the external electron along the field.

We can conclude that in the 
approximation of the infinitely heavy nucleus the number of the ionic bound states is 
infinite, in accordance with the general theorem of \cite{Avron}. 
A quite simple and appealing physical picture of the appearance of such states is that the 
external electron can be attached to the atom with different values of its angular 
momentum along the magnetic field. 
The number of such possible definite values is 
infinite and for each value a different 1D effective potential appears. 
This potential determines the motion of 
the external electron along the field and can bind the electron 
in at least one quantum state. We have labelled these states by the 
integer number $s=0,1,2,\ldots$ (negative of the magnetic quantum number of the 
external electron), and obtained general estimates of the binding energies of 
negatively charged ions. These estimates establish both the dependence of  
the binding energies on the magnetic field strength and on the quantum number $s$:
for $s=0$ the binding energy scales with $B$ as $B^2$ and for $s=1,2,\dots$ the 
scaling low is $-\E_s \propto B^3$; at large $s$ the binding energies behave as 
$-\E_s \propto s^{-3}$. To apply the estimates for $s>0$ to a specific ion one needs 
only to know the polarizability of the corresponding neutral atom in an external 
electric field, while the estimate for $s=0$ involves the whole atomic potential 
acting on a static external electron.

We have applied the estimates obtained to the H$^--$ion, exploiting the known 
polarizability of the hydrogen atom and the potential for the static two-charge 
Coulomb problem \cite{Walles}. We also have verified that these 
estimates are in complete agreement 
with results of the numerical integration of the Schr\"odinger equation for the motion 
of the external electron along the magnetic field. 

The binding energies for the H$^--$ion that we have obtained are indeed very small 
compared to the binding energy of the hydrogen atom. 
At the largest magnetic fields available 
at laboratories now, $B\sim30$~T, the binding energy for the H$^--$ion is 
approximately $2.8\times10^{-3}$~meV for the state with the zero angular 
momentum of the external electron, and it is $\approx 1.4\times10^{-7}$~meV 
for the state with $s=1$. 
The magnetically induced bound states of 
the exotic muon ions possess a much larger binding energy. 
For the same value of the magnetic field strength, 
the muon affinity to the hydrogen atom is 
$0.58$~meV for the state $s=0$ and  
$2.9\times10^{-5}$~meV for the state $s=1$. 

It also looks quite challenging to detect the excited anion states for heavier atoms 
which have polarizabilities significantly larger than the hydrogen atom. For example, 
the polarizability of Cs is, with 2\% accuracy, $\kappa=403$ \cite{Miller}. 
The resulting 
electron affinity to this atom for the state $s=1$ at $B=30$~T is 
$\approx 2\times10^{-4}$~meV. Another challenging example is the anion of Ba, which 
does not exist at $B=0$. The polarizability of Ba, with 8\% accuracy, is 
$\kappa=268$ \cite{Miller}, and the binding energy of its excited anion, 
for the external electron state with $s=1$ at $B=30$~T is $\approx 1.8\times10^{-5}$~meV.
The corresponding states of the exotic ions, Cs$\mu^-$ and Ba$\mu^-$, possess 
the binding energies of $\approx 4\times10^{-2}$~meV and $\approx 4\times10^{-3}$~meV, 
respectively.

Although we cannot directly apply the estimate (\ref{s=0}) for the binding energy of 
the external electron with zero angular momentum to heavier ions, in order to have an 
idea about their order of magnitude we can speculate that these binding energies 
differ from the result (\ref{E_0}) by the factor of $(\kappa/4.5)^2$, 
where $\kappa$ is the polarizability of an atom and $4.5$ is the numerical value of 
the polarizability of the hydrogen atom. In this way we can evaluate the binding energy 
for the state $s=0$ of the external electron of the ion of Cs$^-$, at $B=30$~T, as 
$-\E_0 \sim 20$~meV. For the same state of the ion of Ba$^-$ we obtain, 
at the same field strength, $-\E_0 \sim 10$~meV. 
As we see, the magnetically induced anion bound states for atoms 
with large polarizabilities can very well be detectable in laboratories.

Because of the large extension of the wave function of the loosely bound electron our 
results may also apply to molecules and clusters although in a less quantitative manner. 
As molecules and clusters can exhibit large polarizabilities \cite{Miller} we expect that 
magnetically induced bound states exist for them and possess considerable binding energies. 

One important issue which was not addressed in this paper are 
the finite nucleus mass effects which can affect the ion states significantly.
The underlying physical picture of the finite nuclear motion 
for charged systems was 
discussed in detail in, e.g., \cite{Schmelcher}. Here we only mention that 
the electron states are affected by an oscillating electric field 
introduced in the internal (electronic) system of the ion due to the rotation of the ion as a 
whole over a Landau-like orbit. For a crude estimate, we may assume that due to the  
ion motion the ion's states get the oscillator-like energy excess
$\hbar\Omega N$, where $\Omega=eB/(Mc)$, $M$ is the ion mass and $N=0,1,2,\ldots$. 
Since in the presence of the 
magnetic field the internal (electronic) 
degrees of freedom are inherently coupled to the center of mass ones \cite{Schmelcher}, 
we can expect that similar to the case of the positive He$^+-$ion \cite{Bezchastnov} 
the minimal possible value of $N$ for stable states coincides with $s$. 
In this case, an $s-$state of the 
external electron is still bound if $-\E_s > \hbar\Omega s$, or 
\be
\frac{B}{1~{\rm T}} > \frac{1.58\times 10^4}{A^{1/2} \kappa} \frac{s^{1/2}}{\delta_s}~,
\label{bound_cond}
\ee
where $A$ is the mass number of the ion. According to this criterion, 
for the H$^--$ion the magnetic field strength 
which stabilizes the state $s=1$ against the motional Stark effect, is 
large, $B\approx3.5\times10^3$~T, and cannot be achieved in laboratories. However, 
for heavier ions, the magnetic field which can stabilize them in excited 
states are smaller. For example, for the Cs$^--$ion, such a field is 
$\approx 3.4s^{1/2}\delta_s^{-1}$~T, so the external electron can be bound in a few $s-$states 
at the magnetic fields available at laboratories. We must remark that these estimates 
of the center of mass effects on the ion bound states are very preliminary and based on the 
intuitive picture of the inclusion of the oscillator-like energies associated with the ion 
motion across the magnetic field in the total energy spectrum. 
We plan to perform the corresponding analysis accurately in order to see how the 
nuclear motion 
influences the induced bound states in the magnetic field. But already now we can  
speculate that the finite nuclear mass effects make at least the highly excited $s-$states 
unbound which means that in reality the actual number of bound ion states is 
finite and not infinite.
\vspace*{0.5cm}
\begin{center}
{\bf{Acknowledgements}}
\end{center}
We are grateful to M.V.~Ivanov for providing us with numerically computed wave functions 
for the ground state of an electron moving in the Coulomb fields of two static charges.
One of the authors (VGB) is pleased to acknowledge the Alexander-von-Humboldt 
Stiftung for the Fellowship. Support by the Deutsche Forchungemeinschaft is greatly 
acknowledged. 
\renewcommand{\theequation}{A\arabic{equation}}
\setcounter{equation}{0}
\vspace*{0.5cm}
\begin{center}
{\bf Appendix}
\end{center}

According to Eq.~(\ref{h_00}), the non-adiabatic correction to the potential of interaction 
between the hydrogen atom and an extra electron reads
\be
h_{00}(\vv{r}) = \frac{\hbar^2}{2\me} 
                 \int {\rm d} \vv{r}_1
                 \left| \frac{\partial}{\partial \vv{r}} 
                        \phi_0(\vv{r}_1,\vv{r}) \right|^2~. 
\label{h_hyd_1}
\ee
Here integration is performed over the position $\vv{r}_1$ of the core electron and 
the wave function $\phi_0$ describes the ground state of the atom perturbed by the 
interaction with the static external electron. When the interaction is weak, 
this wave function can be evaluated by the first order perturbation theory,
\be
\phi_0(\vv{r}_1,\vv{r}) = \phi_0^{(0)}(\vv{r}_1) 
                        + \sum_{f \neq 0} \frac{W_{f0}(\vv{r})}{E_0^{(0)}-E_f^{(0)}}
                                          \phi_f^{(0)}(\vv{r}_1)~,
\label{pert_exp}
\ee
where $\phi_0^{(0)}$ and $\phi_f^{(0)}$ are the wave functions of the ground and 
excited states of the non-perturbed atom, respectively, $E_0^{(0)}$ and $E_f^{(0)}$ are 
the corresponding eigenenergies and
\be
W_{f0}(\vv{r}) = \langle \phi_f^{(0)} | W(\vv{r}_1,\vv{r}) | \phi_0^{(0)} \rangle
\ee
are the matrix elements of the perturbation operator
\be
W(\vv{r}_1,\vv{r}) = \frac{e^2}{|\vv{r}_1-\vv{r}|} - \frac{e^2}{r}~.
\label{W_e-hyd}
\ee
Substituting (\ref{pert_exp}) into (\ref{h_hyd_1}) and integrating over $\vv{r}_1$ 
yields, due to the orthogonality of the hydrogenic wave functions,
\be
h_{00} = \frac{\hbar^2}{2\me} 
         \sum_{f \neq 0} \frac{|S_{f0}|^2}{\left( E_0^{(0)}-E_f^{(0)} \right)^2}~,
\label{h_hyd_2}
\ee
where $S = \partial W / \partial \vv{r}$. To simplify the latter expression, 
let us approximate the energy difference in the denominator by the quantity $-$\Ry. 
Then, using the completeness property of the hydrogenic wave functions, we obtain
\be
h_{00} = \frac{\hbar^2}{2\me{\rm Ry}^2} 
         \left[ \left( S^2 \right)_{00} - \left( S_{00} \right)^2 \right]~,
\label{h_hyd_3}
\ee
where $\left( S^2 \right)_{00}$ and $S_{00}$ are the mean values of the operators $S^2$ and 
$S$, respectively, for the ground state of the non-perturbed hydrogen atom. To calculate them, 
it is most convenient to specify the internal atomic coordinates with the quantization axis 
along $\vv{r}$. In these coordinates, assuming $r \gg r_1$, we can replace the operator
(\ref{W_e-hyd}) by the first, dipole, term of its multipole expansion (cf. Eq~(\ref{mult})),  
$W=(e^2/r^2)z_1$. Then we have $S=-(2e^2/r^3)z_1$. For the latter operator, 
because of the $z-$parity of the hydrogenic state, one has $S_{00}=0$ and 
Eq.~(\ref{h_hyd_3}) reduces to
\be
h_{00} = \frac{2\hbar^2e^4\left( z_1^2 \right)_{00}}{\me{\rm Ry}^2r^6}~.
\label{h_hyd_4}
\ee
Substituting there the known value $\left( z_1^2 \right)_{00} = \aB^2$ and introducing 
atomic units, we obtain the estimate of the non-adiabatic correction given 
by Eq.~(\ref{h}), $h_{00}(r)=8/r^6$.

\vspace*{1.0cm} 
{}
\vspace*{0.5cm}
{\bf Figure Captions}

{\bf Figure~1.} The one-particle potential of interaction between the static charge $-e$ 
and the hydrogen atom (solid line) and the non-adiabatic correction to this potential 
for the cases when the charge is associated with the electron (long-dashed line) 
and with the muon (shot-dashed line). 
Dots show the reference data for the potential obtained from \cite{Walles}. 
The part of the solid line connecting the dots is obtained by the spline interpolation, 
at lower $r$ the solid line corresponds to Eq.~(\ref{small_r}) while at large $r$ 
it corresponds to Eq.~(\ref{large_r}).
Open circles show the non-adiabatic corrections obtained numerically, and dashed lines 
correspond to the perturbation estimates. 

{\bf Figure~2.} Binding energies for the states $s=0,1,2,3,4$ (in meV) as functions of 
the magnetic field strength (in Tesla). The conversion rate used is: 
$1$~a.u.~$= 2.35\times10^5$~T. Solid lines show the estimates for the ion 
H$^-$ given by Eqs.~(\ref{E_0}) and (\ref{E_s}), and dots represent the results 
of our numerical treatment. Dashed lines show the corresponding energies for the 
muonic ion, H$\mu^-$.

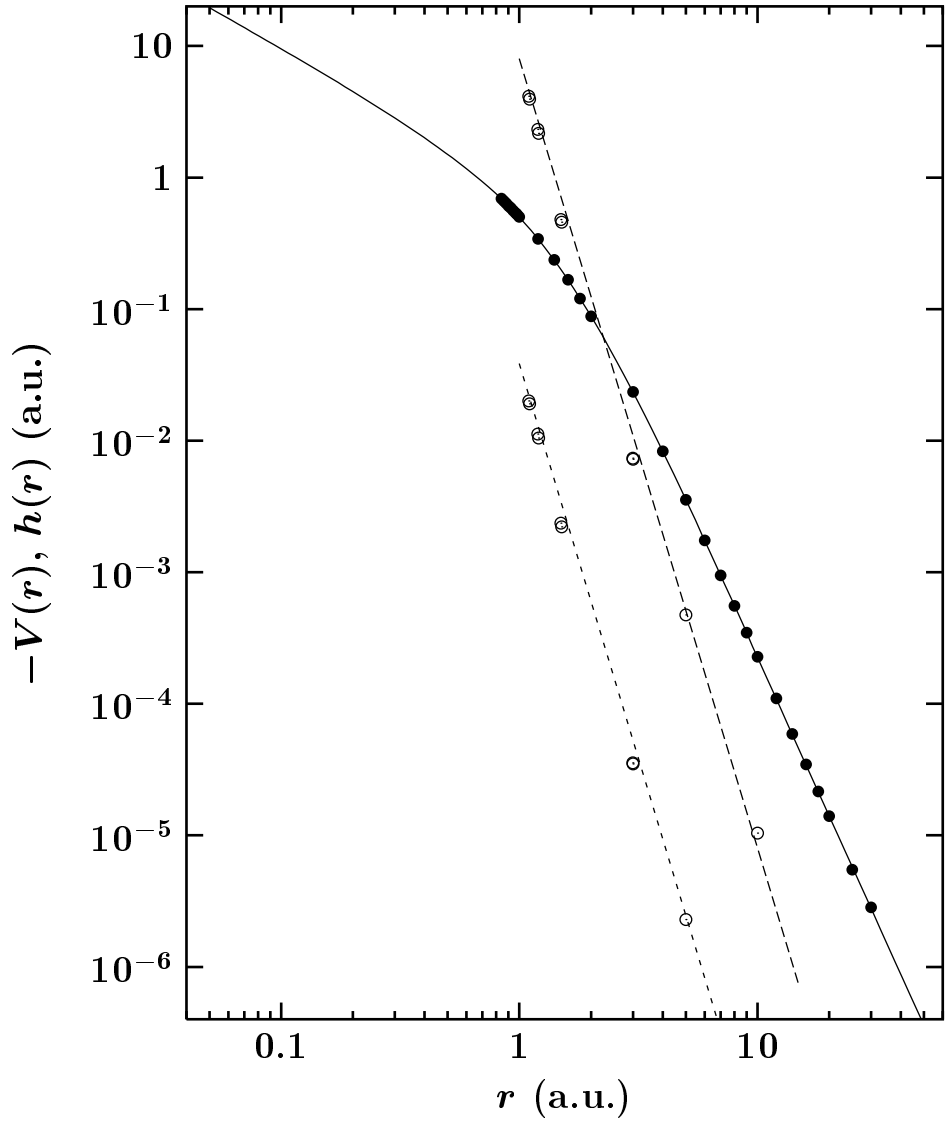
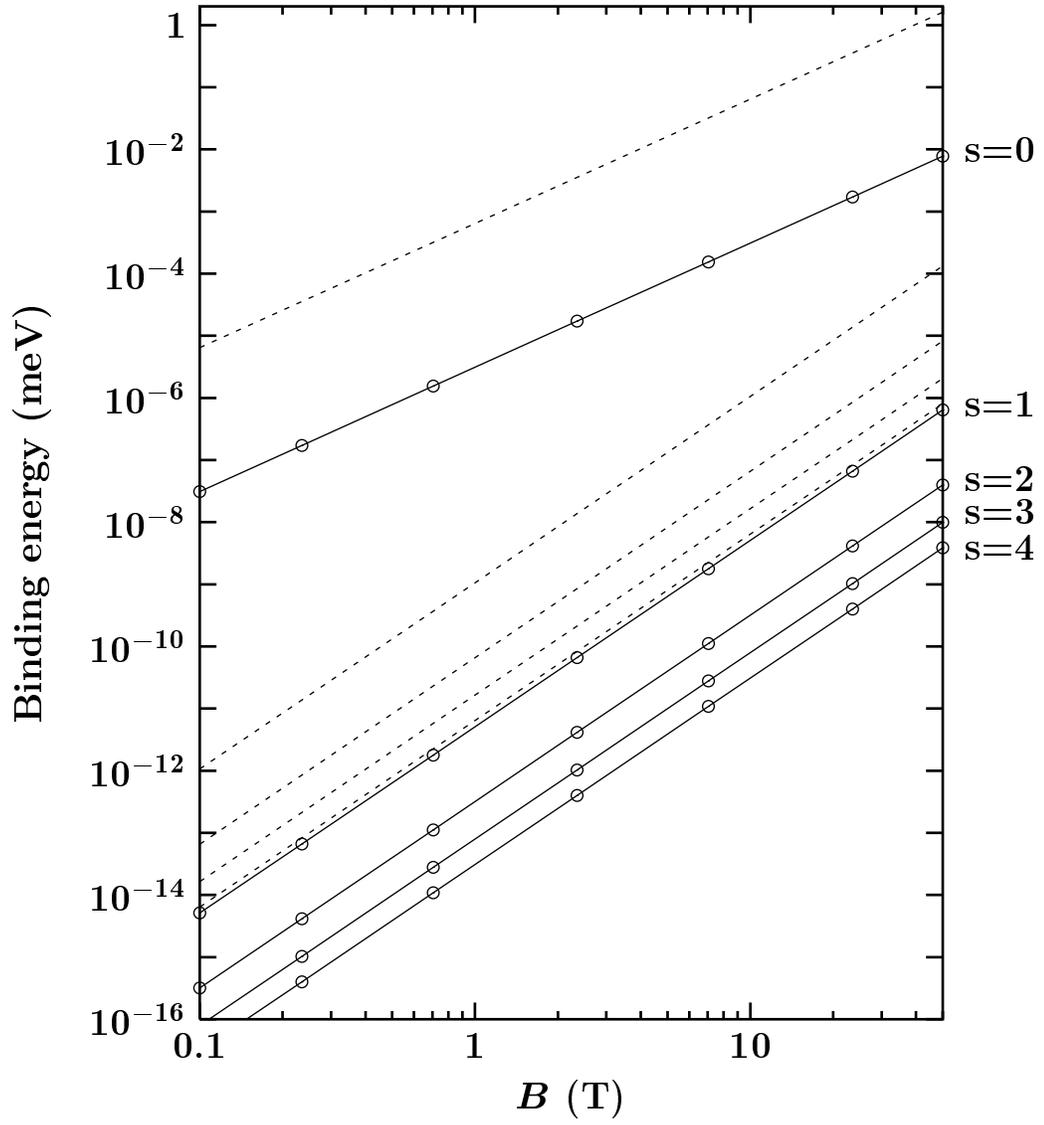
\end{document}